\documentclass[preprint,12pt]{aastex}
\usepackage{psfig}
\input epsf

\def\rmmat#1{{\hbox{\rm #1}}}
\def\rmscr#1{\rmmat{\scriptsize #1}}
\newcommand{\be}{\begin{equation}}
\newcommand{\ee}{\end{equation}}
\newcommand{\ba}{\begin{eqnarray}}
\newcommand{\ea}{\end{eqnarray}}
\def\d{{\rm d}}

\def\dd#1#2{\frac{\d #1}{\d #2}}

\def\eqref#1{Eq.~\ref{eq:#1}}
\begin{document}
\title{Accretion limits the compactness of static stars}
\author{Jeremy S. Heyl\altaffilmark{1}}
\altaffiltext{1}{Chandra Fellow,
Harvard-Smithsonian Center for Astrophysics, 
60 Garden Street, 
Cambridge, MA 02138}

\begin{abstract}
General relativity limits the compactness of static stars.  If the
pressure of the fluid is positive and the density decreases with
distance from the center, the value of the circumferential radius of
the star must be greater than $(9/4) G M/c^2$, or equivalently the
redshift of the surface must be less than two.  If
constraints on the equation of state of the material are relaxed,
general relativity alone does not restrict the redshift of a static
stellar surface.  However, because black hole candidates in the universe
generally accrete material from their environs, the process of
accretion provides upper limits on the redshift of a astrophysical 
black-hole candidates.
\end{abstract}

\section{Introduction}

The Tolman-Oppenheimer-Volkoff \citep{Tolm34,1939PhRv...55..374O}
equations restrict the radius $R$ of a static star of mass $M$
to be greater than $(9/4) G M/c^2$ \citep{Wein72,Shap83}.  There are
two physically motivated assumptions behind this conclusion.  First,
the density of the material must decrease or remain constant with
increasing radius.  Second, the pressure of the fluid must be positive
throughout.  If the properties of the material are less constrained,
this limit may be evaded, but important limits may still be obtained
for astrophysical objects, because these objects interact with their
surroundings.  As a compact object accretes matter, its configuration
must adjust to support the additional material.  This adjustment
cannot occur more quickly than the light-crossing time for the object,
so one can obtain a natural limit to the accretion rate onto a static
star or conversely a limit to the compactness of the star.

The letter examines two stellar models which evade the normal
compactness limit and derives more liberal limits to the value of
$M/R$.  Although the resulting limits are much less constraining than
the standard result, they do have important consequences on the
possible physical processes that could halt the collapse of a
black-hole progenitor.

\section{Calculations}

Outside a spherical static star, Birkhoff's theorem restricts the
spacetime to be Schwarzschild.  Because the focus will be solutions
without horizons, the standard Schwarzschild coordinates are adequate.
The redshift of the stellar surface is given by $1+z = (1 - 2
M/R)^{-1/2}$, in units with $G=c=1$.  An infinite redshift relative to
infinity is achieved where the metric component $g_{00}$ vanishes,
yielding a event horizon and a classical Schwarzschild black hole. 

The ultracompact star may be modelled either as a thin shell with a
radius $R>2 M$ which surrounds vacuum or material at a constant
density with $p=-\rho$.  The properties of the material in the shell
become singular as the thickness of the shell vanishes and can be
determined from the \citet{Isra66,Isra67} matching conditions or
by evaluating Einstein's equation directly.  The first model (case A)
is an extreme counterexample to the assumption that the density must
decrease with radius.  Here the density inside is zero and suddenly
increases at the inner surface of the shell.  The second model (case
B) exhibits negative internal pressure, in fact the maximal negative
pressure which still satisfies the weak-energy condition.

A convenient parameter to characterize the configurations is
$\epsilon=g_{00,\rmscr{surf}}=1-2 M/R$.   In both cases the exterior
metric is given by the Schwarzschild solution.  In case A, the
interior metric is
\begin{equation}
ds^2 = \left ( 1 - \frac{2 M}{R} \right ) dt^2 - dr^2 
	- r^2 \left ( d\theta^2 + \sin^2\theta d \phi^2 \right ).
\end{equation}
A simple redefinition of the time coordinate reveals that the interior
spacetime is flat.  The gravitational redshift relative to infinity is
constant inside the shell.

In case B, the metric is slightly more complicated
\begin{equation}
ds^2 = \left ( 1 - \frac{2 M(r)}{r} \right ) dt^2 - \left ( 1 -
\frac{2 M(r)}{r} \right )^{-1} dr^2 - r^2 \d \Omega^2
\end{equation}
where $M(r) \propto r^3$ is the mass enclosed within the radius $r$.
Inside the star, the redshift relative to infinity decreases from its
value at the surface to zero at the center of the star.  The
configuration discussed in detail by 
\citet{2001gr.qc.....9035M} falls into this latter class.

An important characteristic of these solutions is the coordinate time
(the time measured at infinity) for light to go from the surface of
the star to the center.  This is the shortest time that the star can
adjust its configuration without violating causality.  For the mass
shell (case A), $\tau = \epsilon^{-1/2} R$.  If the accretion is
perfectly sphereically symmetric, the shell could in principle adjust
on a faster timescale of $\tau \sim \epsilon^{1/2} R$, the time for a
signal to propagate from the surface of the shell to where the Schwarzschild
radius would be.  However, the accretion is unlikely to be perfectly 
symmetric, so different parts of the shell would have to be in causal contact
for the entire shell to adjust, and the longer timescale is appropriate.

The constant density (case B), $p=-\rho$ sphere has a crossing time of
\begin{equation}
\tau =  \frac{\tanh^{-1} \sqrt{1-\epsilon}}{\sqrt{1-\epsilon}} R.
\end{equation}
As $\epsilon$ approaches zero, this expression diverges as $\ln
\epsilon$, more gently than the mass-shell solution.  From the
structure of the Tolman-Oppenheimer-Volkoff
\citep{Tolm34,1939PhRv...55..374O} equations, it is apparent that this
solution has the minimal light-crossing time for a configuration with
a given surface redshift, while satisfying the weak-energy condition.

As material falls onto or through the surface of the star, the value
of $\epsilon$ will change.  The rate of change in $\epsilon$ is
proportional to the rate of increase of the gravitational mass of the
star
\begin{equation}
\dd{\epsilon}{t} = -\frac{2}{R} \dd{M}{t} 
\left (1 - \frac{1}{3} \frac{\bar \rho}{\rho} \right )
\end{equation}
where $\rho$ is the density at which the material accumulates and
$\bar \rho$ is the mean density of the star defined by $M/(4/3 \pi
R^3)$.  If the accreted material simply sinks through the surface and does
not accumulate there, the effective ratio ${\bar \rho}/\rho$ vanishes.  In
the subsequent constraints $\rho$ is taken to equal ${\bar \rho}$.

For a particular configuration, if $\epsilon + \tau d\epsilon / d t$
is negative, a trapped surface will form at the stellar surface.  In
case A the star must collapse to form a singularity \citep{Hawk70}.  In
case B, the formation of a singularity may be avoided
\citep{1990NuPhB.339..417F}.  In both cases, the object will appear
like a classical black hole to outside observers.  Essentially, for
sufficiently fast accretion rates, the star cannot adjust its
configuration quickly enough to avoid the formation of a horizon.

\section{Astrophysical Implications}

\subsection{Eddington-Limited Accretion}
\label{sec:edd}

Black-hole candidates have been discovered with masses ranging from a
few solar masses to a few billion solar masses.  Furthermore,
black-hole candidates across this entire range of masses have been
observed to accrete up to the Eddington accretion rate.  At this rate,
the outgoing photon flux is sufficient to quench additional accretion.
In geometrized units it is given by ${\dot M}=3 \times 10^{-22}
\gamma^{-1} M/{\rm M}_\odot$ where $\gamma$ is the efficiency
of the energy release $\sim 0.1$.

Because only a small fraction of the rest-mass energy of the infalling
material is radiated to infinity, the gravitational mass of the
black-hole candidate must increase, specifically, 
$d M/d t=(1-\gamma) {\dot M}$.  At the Eddington accretion rate,
$d\epsilon /d t=1.6 \times 10^{-27} (1-\gamma)/\gamma$~cm$^{-1}$.

This yields an upper limit on the value of $\epsilon$ which is a
function of the mass of the black-hole candidate and the spacetime in
its interior.  For case A (the mass shell),
\begin{equation}
\epsilon > 
	2.5 \times 10^{-14} \left ( \frac{M}{{\rm M}_\odot} \right )^{2/3}
\end{equation}
and for case B (the constant density interior)
\begin{equation}
\epsilon > 10^{-18}  \frac{M}{{\rm M}_\odot} 
\left [ 1 - \frac{1}{50} \ln \left ( \frac{M}{{\rm M}_\odot} \right ) \right ]
\end{equation}
where $\gamma=0.1$ and $\rho={\bar \rho}$.  Although these limits may
not appear particularly stringent, the proper distance, $l$, that the
surface of the star must move to lie within its horizon is much
greater than the Planck length.  It is typically $1 (M/{\rm
M_\odot})^{4/3}$~mm for the mass shell and $6 (M/{\rm M_\odot})^{3/2}$
$\mu$m for the negative-pressure star.

\subsection{Formative Accretion}

Although the details of the formation of supermassive black holes are
hazy, black holes of stellar mass form in supernovae.  Numerical
simulations \citep{1999ApJ...524..262M} indicate that during a
supernova that results in a black hole that the mass accretion rate
can exceed $0.1 \rmmat{M}_\odot \rmmat{s}^{-1}$ or $5\times 10^{-7}$
in geometrized units.  This is fourteen orders of magnitude larger
than the Eddington accretion rate for a ten-solar-mass object.  The
system evades the Eddington limit because the accreting material is
sufficiently hot that neutrinos rather than photons are the dominant
radiative channel \citep{2001ApJ...557..949N}.  In both cases, if the
objects formed in supernova are not black holes but static stars, the
limiting value of $\epsilon$ is much larger than for the Eddington
accretion rate achieved later in the object's lifetime.  For case A
$\epsilon > 7 \times 10^{-5}$, and for case B
$\epsilon > 5 \times 10^{-6}$.

\subsection{Mergers}

Supermassive black holes grow through a succession of mergers as their
parent galaxies merge \citep{1980Natur.287..307B,2001ApJ...558..535M}.
Unlike accretion during the formation of a stellar-mass black hole,
this process is neither quasistatic nor quasispherical.  However, the
general arguments do apply in this case.  Specifically, to prevent the
formation of a horizon, the two objects must adjust their
configurations during the process of the merger, and this requires at
least a light-crossing time.  Because the final plunge of the merging
objects is approximately head-on at nearly the speed of light
\citep{2000PhRvL..85.5496B}, the effective accretion rate is $\sim \d
M/\d t \approx M_2/(2 M_1)$, where $M_1$ and $M_2$ are the masses of
the primary and secondary respectively.  During the final stage of an
equal-mass merger, using the same formulae as in the earlier
calculations, unless the radius of each object is larger than $\approx
4 M$, a horizon will form around the coalescing pair in less than
internal dynamical time.  Detailed simulations of merging black holes
support this conclusion \citep{2000PhRvL..85.5496B,2001PhRvL..87A1103A}.

\section{Discussion}

If the mass of a protoneutron star exceeds several solar masses, is
collapse to a black hole inevitable as the star cools?  Or to ask in
another way, must black-hole canditates with masses greater than
several solar masses indeed be black holes?  As a  black-hole or
neutron-star progenitor collapses, the comoving density of the material
increases.  If the mass of the object is less three solar-masses or
so, the collapse will halt and a neutron star will form.  The typical
density of a neutron star is $\sim 10^{15}$~g/cm$^3$ \citep{Shap83}.
The mean density of a ten-solar-mass collapsing star as it passes
through its Schwarzschild radius is an order of magnitude smaller than
this.  The existence of neutron stars indicates that nothing
extraordinary happens to the pressure of the material in the
collapsing star as its surface passes through its Schwarzchild radius.

\citet{2002A&A...396L..31A} argue that observations of material
orbiting and falling toward black-hole candidates cannot determine
with certainty that the central object is indeed a black hole.
However, if general relativity is correct, the process of accretion in
itself provides important limits on what the central object could be.
Specifically, if our understanding of how stellar-mass black holes
form is correct, the redshift of the surface of a black-hole candidate
must not exceed $10^3$.  Admittedly radiation from such a surface
would be difficult to detect, but this maximal redshift is much less
than the value $10^{19}$ considered by \citet{2002A&A...396L..31A}.  

Within classical general relativity, nothing marks the moment the
stellar surface passes through its Schwarzschild radius.  Quantum
mechanically the Schwarzschild radius may be singular due the the back
reaction of the Hawking radiation on the spacetime
\citep{1984PhRvD..29..628Z,1998NuPhS..68..174T}.  Mazur and Mottola
\citep{2001gr.qc.....9035M} argue that this backreaction may result in
a phase transition resulting in the formation of a static object whose
radius is slightly larger than $2 M$.  In the particular model of
Mazur and Mottola, the interior of the object consists of a fluid with
$p=-\rho$, so the compactness of the configuration must be limited,
otherwise the object will be unstable as it accretes.  Using the most
convervative estimate for the maximum value of $M/R$ discussed in
\S~\ref{sec:edd}, the local energy-density in the quantum-backreaction
field is only $\sim (0.3 (M/{\rm M_\odot})^{-3/2} \rmmat{ev})^4$.  If
the required energy density to induce the phase transition is larger
than this (as one would expect), the static star that forms would
collapse if it accreted material at the Eddington rate.

The dynamics of accretion onto black-hole candidates combined with the
condition that any static star must adjust in a causal manner to
changes in its parameters requires either that the redshift of the
surfaces of black-hole candidates be limited, that they are not static
or consist of material which does not satisfy the weak-energy
condition.  Although the limit on the redshift of the surfaces is not
stringent enough to allow detection of the surfaces of stellar-mass
black holes (with $z<10^3$), it does restrict the physical processes
that could halt the collapse of the black hole progenitor to low
energies and densities.  If supermassive black holes mass grow by
mergers, the redshift of their surfaces must be less than 0.2 or they
must lack surfaces entirely.  Such a surface could easily be detected,
so it is unlikely that supermassive black holes have static surfaces.

\acknowledgments

I have been supported by the Chandra Postdoctoral Fellowship Award \#
PF0-10015 issued by the Chandra X-ray Observatory Center, which is
operated by the Smithsonian Astrophysical Observatory for and on
behalf of NASA under contract NAS8-39073.

\bibliographystyle{apj}
\bibliography{ns,mine,gr}
\end{document}